# Direct synthesis and chemical vapor deposition of 2D carbide and nitride MXenes

Di Wang[1], Chenkun Zhou[1], Alexander S. Filatov[1], Wooje Cho[1], Francisco Lagunas[2], Mingzhan Wang,[3] Suriyanarayanan Vaikuntanathan[1], Chong Liu,[3] Robert F. Klie[2], Dmitri V. Talapin[1,3,4*]

[1]*Department of Chemistry and James Franck Institute, University of Chicago, Chicago, Illinois 60637, United States*

[2]*Department of Physics, University of Illinois Chicago, Chicago, Illinois 60607, United States*

[3]*Pritzker School of Molecular Engineering, University of Chicago, Chicago, Illinois 60637, United States*

[4]*Center for Nanoscale Materials, Argonne National Laboratory, Argonne, Illinois 60439, United States*

*Correspondence to: dvtalapin@uchicago.edu

**Summary:** Various MXenes can be synthesized directly by solid-state reactions and, by chemical vapor deposition, in the form of vertical MXene carpets optimal for efficient ion intercalation.

## ABSTRACT

Two-dimensional (2D) transition metal carbides and nitrides (MXenes) are a large family of materials actively studied for various applications, especially in the field of energy storage. To date, MXenes are commonly synthesized by etching the layered ternary compounds, MAX phases. Here we demonstrate a direct synthetic route for scalable and atom-economic synthesis of MXenes, including phases that have not been synthesized from MAX phases, by the reactions of metals and metal halides with graphite, methane or nitrogen. These directly synthesized MXenes showed excellent energy storage capacity for Li-ion intercalation. The direct synthesis enables chemical vapor deposition (CVD) growth of MXene carpets and complex spherulite-like morphologies. The latter form in a process resembling the evolution of cellular membranes during endocytosis.

**Main Text**

MXenes, where M stands for an early transition metal (Ti, V, Nb, Mo, *etc*.) and X is C or N, are a large family of two-dimensional (2D) transition metal carbides and nitrides. Since the discovery of $Ti_3C_2T_x$ (T = O, OH, and F) in 2011 (*1*), MXenes have been commonly synthesized from crystalline MAX phases (A being Al, Si, Zn, *etc*.) by selective etching of A atoms with HF-containing solutions (*1-3*) or Lewis acidic molten salts (*4, 5*), followed by the delamination of the MXene sheets (*6*). Interest in MXenes continues to grow due to their prospects for applications in energy storage (*7, 8*), electromagnetic interference (EMI) shielding (*9, 10*), transparent conductive layers (*11*), superconductivity (*5*), and catalysis (*12*). Moreover, the aforementioned T components in MXenes can be replaced with covalently-bonded surface groups, including organic molecules, either during etching of the MAX phase (*4, 13*), or by post-synthetic modifications of surface groups (*5*). As such, opportunities are available to combine the benefits of 2D MXenes, such as low metal diffusion barrier (*14*), excellent electrical and thermal conductivity (*3*), with the nearly endless tailorability of molecular surface groups.

Traditional preparations of MXenes by high-temperature synthesis and chemical etching of MAX (*15*) or non-MAX (*16, 17*) phases require high energy consumption, show poor atom economy, and use large amounts of hazardous hydrofluoric acid or Lewis acidic molten salts. The development of direct synthetic methods would greatly facilitate practical applications of the rapidly developing family of functional MXenes. An ideal approach would involve a reaction of inexpensive precursors into MXenes bypassing intermediate MAX phases. In 2019, Druffel *et al*. reported the synthesis of $Y_2CF_2$ with MXene-like structure by the solid-state reaction between $YF_3$, Y metal, and graphite (*18*), based on the previously reported synthesis of Y, Sc, and Zr metal carbide halides by Corbett *et al*. in 1986 (*19*).

Among about a hundred known MXene structures, titanium MXenes show some of the best combinations of physical and chemical properties (*20*) relevant for a variety of applications (*21*). Here we show that $Ti_2CCl_2$ and $Ti_2NCl_2$ MXenes can be directly synthesized from Ti metal, titanium chlorides ($TiCl_3$ or $TiCl_4$), and various carbon or nitrogen sources (graphite, $CH_4$, or $N_2$). These directly synthesized MXenes (denoted as ds-MXenes) can be delaminated, and their surface groups can be replaced with other molecules by nucleophilic substitution or completely removed by reductive elimination reactions (*5*). Besides convenience and scalability, the direct synthesis routes offer new synthetic modalities not compatible with traditional MAX etching methods. For example, we demonstrate chemical vapor deposition (CVD) synthesis of extended carpets of $Ti_2CCl_2$, $Ti_2NCl_2$, $Zr_2CCl_2$, and $Zr_2CBr_2$ MXene sheets oriented perpendicular to the substrate. Such orientations make MXene surfaces easily accessible for ion intercalation (*7, 22*) and (electro)chemical transformations (*23, 24*) by exposing edge sites with high catalytic activity (*25, 26*).

A schematic of the new ds-$Ti_2CCl_2$ synthesis by high-temperature reaction between Ti, graphite and $TiCl_4$ is shown in Fig. 1A. This reaction can be carried out on a multigram scale (Fig. S1) and should be easily amenable to further scaling. Powder X-ray diffraction (XRD) and structural analysis by Rietveld refinement of the as-synthesized reaction products (Fig. 1B) show the presence of a $Ti_2CCl_2$ MXene phase with the lattice parameters $a = 3.2284(2)$ Å and $c = 8.6969(1)$ Å, which are close to the values reported for $Ti_2CCl_2$ MXene synthesized by etching of $Ti_2AlC$ MAX phase with Lewis acidic molten salt (referred to as MS-MXenes) (*5*). Cubic $TiC_x$ ($x = 0.5$-$1$) was often present as a byproduct but could be efficiently removed by its precipitation from nonaqueous dispersions of the raw product prepared, e.g., by ultrasonic dispersion in propylene carbonate (PC) or by delamination of ds-$Ti_2CCl_2$ with *n*-butyllithium (*n*-BuLi) (Figs. 1C, S2 and Supplementary Note 1).

The formation of ds-$Ti_2CCl_2$ MXene was observed beginning at around 850°C, and the yield of MXene was maximal at 950°C (Supplementary Note 2). $TiC_x$ became the dominant reaction product at temperatures higher than 1000°C. At 950°C, the formation of $Ti_2CCl_2$ phase was observed after two hours, and the ratio between ds-$Ti_2CCl_2$ and $TiC_x$ products did not change significantly after increasing reaction time from 2 hours to 10 days at this temperature. This naturally raises a question as to whether MXene is the kinetic or thermodynamic product of the reaction described in Fig. 1A. We noticed that MXene phase did not form when we attempted to react $TiC_x$ with Ti and $TiCl_3$ or $TiCl_4$. On the other hand, a prolonged heating of purified MS-$Ti_2CCl_2$ at 950°C resulted in a partial conversion into $TiC_x$ (Fig. S3B). We can conclude from these observations that $Ti_2CCl_2$ is a kinetically favored phase forming in competition with $TiC_x$.

The XRD patterns of ds-$Ti_2CCl_2$ synthesized from $TiCl_3$ or $TiCl_4$ are very similar (Supplementary Note 2), as are scanning electron microscopy (SEM) images of the products' morphology, represented by large MXene stacks (Figs. 1D, E, S4, S5). A high-resolution high-angle annular dark field (HAADF) scanning transmission electron microscopy (STEM) image of ds-Ti$_2$CCl$_2$ MXene oriented along the [2$\bar{1}\bar{1}$0] zone axis and its corresponding electron energy loss spectroscopy (EELS) elemental maps are shown in Fig. 1F and Fig. S6, respectively. The center-to-center distance between MXene sheets calculated from HAADF images is about 0.88 nm (Fig. S7A), in agreement with the value of 0.87 nm measured from XRD. ds-Ti$_2$CCl$_2$ MXene sheets showed an atomic ratio of Ti to Cl very close to an ideal 1:1 stoichiometry, suggesting full coverage of MXene surfaces with Cl is achieved (Fig. S8). In comparison, the MXenes synthesized by the traditional MAX-exfoliation route often are significantly deficient in surface coverage, with a typical stoichiometry of Ti$_2$CCl$_{1.5-1.7}$ (*5*). All these structural features, together with X-ray photoelectron spectroscopy (XPS) (Figs. S9A, 10) and assessment of crystal quality from linewidths in Raman spectra (Fig. S12), confirm the structural perfection of our ds-Ti$_2$CCl$_2$ MXene product.

As-synthesized ds-Ti$_2$CCl$_2$ MXene stacks can be delaminated and solution-processed as individual 2D monolayers (Figs. 1C, S13). For delamination, multilayer MXene is first intercalated with Li$^+$ by treatment with *n*-BuLi in hexanes solution (Fig. S2A) (*5, 27*), then shaken with polar solvents such as *N*-methylformamide (NMF) to form a stable suspension of delaminated 2D sheets (Fig. S14). Insoluble TiC$_x$ byproducts can be selectively precipitated by a mild centrifugation at 240 g for 15 min (Fig. S2B). In delaminated ds-Ti$_2$CCl$_2$, the (0001) diffraction peak shifts to a lower 2θ angle of 7.02°, corresponding to the enlarged *d*-spacing value of 12.54 Å, from the original 8.70 Å. A similar *d*-spacing expansion was found in delaminated Ti$_3$C$_2$Cl$_2$ MXenes (from 11.08 Å to 14.96 Å) (*5*).

MXenes are known for excellent pseudocapacitive energy storage properties originating from efficient ion intercalation between 2D layers, with Ti$_2$CT$_x$ MXenes showing some of the highest predicted and experimentally observed capacities among all studied MXene materials (*20, 28*). We therefore investigated the Li-ion storage properties of electrodes prepared from ds-Ti$_2$CCl$_2$ (see Supplementary Note 3 for details). Fig. S15 shows a typical cyclic voltammetry (CV) profile recorded at a potential scan rate of 0.5 mV s$^{-1}$ within the window of electrochemical potentials from 0.2 to 3 V vs. Li$^+$/Li. The specific capacitance of ds-MXene electrode was 341 F g$^{-1}$, in good agreement with previously reported data for MS-Ti$_2$CCl$_x$ MXene prepared by MAX phase exfoliation synthetic route (*28*). The rectangular CV profile without redox peaks suggests a pseudocapacitive energy storage mechanism for delaminated MXenes (*29*), which is further supported by the consistency of the rectangular CV profiles recorded with different negative cut-off potentials (Fig. 1G). A maximum capacity of 286 mAh g$^{-1}$ was recorded at a specific current of 0.1 A g$^{-1}$ within 0.1 to 3 V (Fig. S16A), which is slightly higher than previously reported data for the optimized performance of MS-Ti$_2$CCl$_x$ MXene (*28*). These electrochemical studies further confirm the excellent quality of ds-Ti$_2$CCl$_2$ MXene.

Excellent electrochemical energy storage characteristics of MXenes stem from the combination of large surface-to-volume ratio and high electrical conductivity. However, restacking of MXene sheets can reduce the surface area easily accessible for intercalating ions, which is a well-known problem of 2D materials (*30*). Further expansion of synthetic methodology may allow arrangements of MXene sheets with more easily accessible surfaces and exposed catalytically active edge sites. As a demonstration, we introduce the direct synthesis of MXenes through CVD and show that CVD offers a route to completely new morphologies of MXenes. Although transition metal carbides and nitrides such as Mo$_2$C, WC, and Mo$_2$N can be grown by CVD (*31, 32*), such synthetic option has not been previously available for MXenes. We now succeed in growing MXenes by CVD on a Ti surface with a CH$_4$ and TiCl$_4$ gas mixture diluted in Ar (Fig. 2A). After the exposure of Ti foil to TiCl$_4$ and CH$_4$ vapor at 950ºC for 15 min, the as-synthesized product (denoted as CVD-Ti$_2$CCl$_2$) was characterized by XRD (Fig. 3B). According to the Rietveld refinement, the lattice parameters *a* = 3.2225(2) Å and *c* = 8.7658(8) Å match well the reported values for Ti$_2$CCl$_2$ MXene (*5*). Raman spectra (Fig. 3C) also confirm the purity of Ti$_2$CCl$_2$ MXene. High resolution STEM-EELS (Fig. S17) and EDX analysis (Fig. S18) confirm the crystallinity and ideal stoichiometry of CVD-Ti$_2$CCl$_2$. The center-to-center interlayer distance of 0.88 nm calculated from STEM images was typical for Ti$_2$CCl$_2$ MXenes (Fig. S7B). SEM images showed a substrate fully covered with a wrinkled layer of CVD-Ti$_2$CCl$_2$ (Fig. 2D). A dense carpet of Ti$_2$CCl$_2$ MXene sheets grown perpendicular to the substrate would be difficult to achieve for traditionally-synthesized MXenes. This morphology appears particularly promising for efficient ion intercalation and fast charging/discharging cycles, see Supplementary Note 3 (*7, 22*).

CVD growth of MXenes involves the reaction of gaseous reagents with the titanium surface, as in the preparation of carbon nanotubes by the root-growth mechanism (*33*). As the thickness of growing MXene carpet increases, the diffusion of gaseous reagents toward the reaction zone (Fig.

3A) slows down, and the growth of the MXene carpet is expected to be self-limiting. However, we observed emergence of a new growth mechanism that bypassed this kinetic bottleneck (Fig. 3B). The uniform growth of MXene carpet (Fig. 2D) was followed by the formation of "bulges" (Fig. 3C) that further evolved into MXene "vesicles" (Fig. 3D). Next, these "vesicles" detached from the substrate and can be collected and imaged by SEM and TEM (Figs. 3D-F and Supplementary Note 4). The process repeats itself, enabling continuous synthesis of MXenes. The internal structure of CVD-MXene "vesicles" was composed of individual $Ti_2CCl_2$ sheets radiating from the center and oriented normal to the surface (Figs. 3G, H, S19). Imaging of a fragmented "vesicle" (Fig. 3E) and individual "vesicles" dissected with a focused ion beam (FIB, Fig. S20) reveal a small void at the "vesicle" centers.

The complexity of this hierarchical organization of CVD-$Ti_2CCl_2$ "vesicles" is unusual for inorganic solids. In Supplementary Note 4, we discuss their possible growth mechanism, inspired by the non-equilibrium evolution of cell and organelle membranes. The MXene carpet formed at an early stage of CVD growth (Fig. 2C) can be approximated as an elastic sheet with the equilibrium energy defined through the surface area $S$, surface tension $\gamma$, local curvature, and bending rigidity $\kappa$. If $\gamma, \kappa > 0$, the sheet naturally prefers equilibrium flat geometry (*34*). However, when new material is constantly added to the sheet, the standard equilibrium description fails to predict its shape and stability (*35*). When elastic membranes are forced to grow, their evolution can be modeled in terms of an effective decreased surface tension, $\gamma_{eff} = \gamma - \alpha \dot{m}$, where $\dot{m}$ is the rate at which material is added to the sheet and $\alpha$ is a phenomenological constant (*36, 37*). If the growth rate exceeds a critical threshold, $\dot{m}_c = \gamma/\alpha$, the effective surface tension takes negative values. A negative surface tension then implies that certain fluctuations of the elastic sheet grow rapidly resulting in an instability. Van der Waals bonded 2D MXene sheets can efficiently slide against each other, creating only a small elastic penalty for the formation of buckled and curved geometries. If the sheet is just loosely connected to underlying substrate, these deformations can collapse into spherical "vesicles" refreshing substrate for further growth, as schematically shown in Fig. 3B. Such evolution of MXene carpets during CVD growth conceptually resembles the dynamics of cell membranes during endocytosis (*38*). A recent theoretical work illustrated how negative surface tension brought about by growth could lead to a variety of non-trivial geometries similar to our experimentally observed MXene "vesicles" (*36*). Various supporting evidence was found for the above sequence of growth stages (see Supplementary Note 5), but further in-depth studies will be needed to fully rationalize this, to the best of our knowledge, unprecedented mechanism of CVD growth of 2D materials.

Direct CVD synthesis can be employed to produce MXenes that have not been previously prepared by the etching of MAX phases. For example, $Zr_2CCl_2$ and $Zr_2CBr_2$ MXenes were synthesized by exposing a Zr foil to $CH_4$ and $ZrCl_4$ or $ZrBr_4$ vapor at 975ºC. These two zirconium MXenes appeared in the same general morphology as the titanium MXenes, adopting vertically aligned carpet-like structure on the surface of the Zr foil (Figs. S21).

Arguably the most intriguing product of the direct synthesis was phase-pure nitride $Ti_2NCl_2$ MXene formed via the reaction of Ti foil with $TiCl_4$ and $N_2$ above 600ºC (Figs. 2B, C, E, S22). To the best of our knowledge, neither this reaction nor this MXene phase have been reported previously, but it has been predicted for nitride MXenes to have a variety of attractive properties, including ferromagnetism and higher conductivity as compared to carbide MXenes (*39*). To date, only a few nitride MXenes have been synthesized, and experimental realization of chlorine terminated nitride MXenes have not been achieved. Our CVD method, using $N_2$ as the nitrogen source, further proves the versatility of bottom-up MXene syntheses. These reactions can be important beyond MXenes synthesis. Given that $TiCl_4$ plays the key role in Ti metallurgy (Kroll process) and in synthesis of $TiO_2$ from titanium ores (chloride process), both being on the millions of tons annually, the above reactions may create interesting opportunities, e.g., for nitrogen fixation as a side process in conventional $TiO_2$ synthesis.

**Acknowledgements**

We thank Yu Han and Gangbin Yan for helpful discussions of electrochemical measurements and Dr. Gerard Olack for help with SEM data analysis. We are also grateful to Andrew Nelson for a critical reading and editing of the manuscript.

**Funding:** The work on direct MXene synthesis was supported by the National Science Foundation under award number DMR-2004880, and CVD synthesis was supported by the Department of Defense Air Force Office of Scientific Research under grant number FA9550-18-1-0099. Electrochemical studies were supported by the Advanced Materials for Energy-Water Systems (AMEWS) Center, an Energy Frontier Research Center funded by the U.S. Department of Energy, Office of Science, BES. W.C. and S.S. were supported by the University of Chicago Materials Research Science and Engineering Center, which is funded by the National Science Foundation under award number DMR-2011854. S. V. acknowledges support from the National Science Foundation under Grant No. DMR-1848306. F.L. and R.F.K. at UIC were supported by a grant from the National Science Foundation (NSF-DMR 1831406). Acquisition of UIC JEOL ARM200CF was supported by an MRI-R2 grant from the National Science Foundation (DMR-0959470). The Gatan Continuum GIF acquisition at UIC was supported by an MRI grant from the National Science Foundation (DMR-1626065). The work used resources of the Center for Nanoscale Materials, a U.S. Department of Energy (DOE) Office of Science User Facility operated for the DOE Office of Science by Argonne National Laboratory under Contract No. DE-AC02-06CH11357.

**Author Contributions:** D.W. performed and designed the experiments, analyzed data, and co-wrote the paper. C.Z. carried out Raman measurements and data analysis. A.S.F. contributed to X-ray measurements and data analysis. W.C. contributed to TEM analysis of delaminated MXene and building the CVD system. F.L. and R.F.K. performed high resolution STEM studies and image analysis. M.W. and C.L. contributed to the electrochemistry measurements and data analysis. S.V. performed simulations and interpretation of the morphology of CVD-MXenes. D.V.T. conceived and designed experiments and simulations, analyzed data, co-wrote the paper, and supervised the project. All authors discussed the results and commented on the manuscript.

**Competing interests:** None declared.

**Data and materials availability:** All data needed to evaluate the conclusions in the paper are present in the paper or the Supplementary Materials. The samples can be provided by the authors upon reasonable request under a materials transfer agreement with the university. Correspondence and requests for materials should be addressed to D.V.T. (dvtalapin@uchicago.edu).

**Figure Captions**

**Figure 1. Direct synthesis and characterization of ds-Ti$_2$CCl$_2$ MXene**. (**A**) Schematic diagram of the synthesis. (**B**) XRD pattern and Rietveld refinement of ds-Ti$_2$CCl$_2$ prepared by reacting Ti, graphite and TiCl$_4$ at 950°C. (**C**) XRD patterns of dispersible delaminated and sonicated ds-Ti$_2$CCl$_2$ MXenes. Inset: Colloidal solution of the delaminated ds-Ti$_2$CCl$_2$. (**D**) SEM image and (**E**) EDX elemental mapping of a ds-Ti$_2$CCl$_2$ stack. (**F**) High resolution HAADF image representing the layered structure of ds-Ti$_2$CCl$_2$. (**G**) CV profiles of delaminated ds-Ti$_2$CCl$_2$ with various negative cut-off potentials.

**Figure 2. CVD growth of MXenes.** (**A**) Schematic diagram of the CVD reactions. (**B**) XRD patterns and Rietveld refinement for CVD-Ti$_2$CCl$_2$ and CVD-Ti$_2$NCl$_2$. (**C**) Raman spectra of CVD-Ti$_2$CCl$_2$ and CVD-Ti$_2$NCl$_2$ MXenes in comparison to that of a traditional MS-Ti$_2$CCl$_2$ MXene, which was synthesized by etching Ti$_2$AlC MAX phase with CdCl$_2$ molten salt. (**D**) Frontal and cross-sectional SEM images of CVD-Ti$_2$CCl$_2$. (**E**) High resolution HAADF images and EELS elemental mapping of CVD-Ti$_2$NCl$_2$.

**Fig. 3. Morphologies of CVD-Ti$_2$CCl$_2$.** Schematic diagram illustrating the (**A**) reaction zone and (**B**) proposed buckling mechanism of CVD-Ti$_2$CCl$_2$ through which microspheres are formed. SEM images show that morphology of CVD-Ti$_2$CCl$_2$ can be varied by tuning reaction conditions: (**C**) Microspheres growing on carpets, (**D**) individual microspheres, and (**E**) a fragmented microsphere showing a hollow center. (**F, G, H**) STEM analysis further proves that vertically aligned MXene sheets constitute the microspheres, while a void is left at the center.

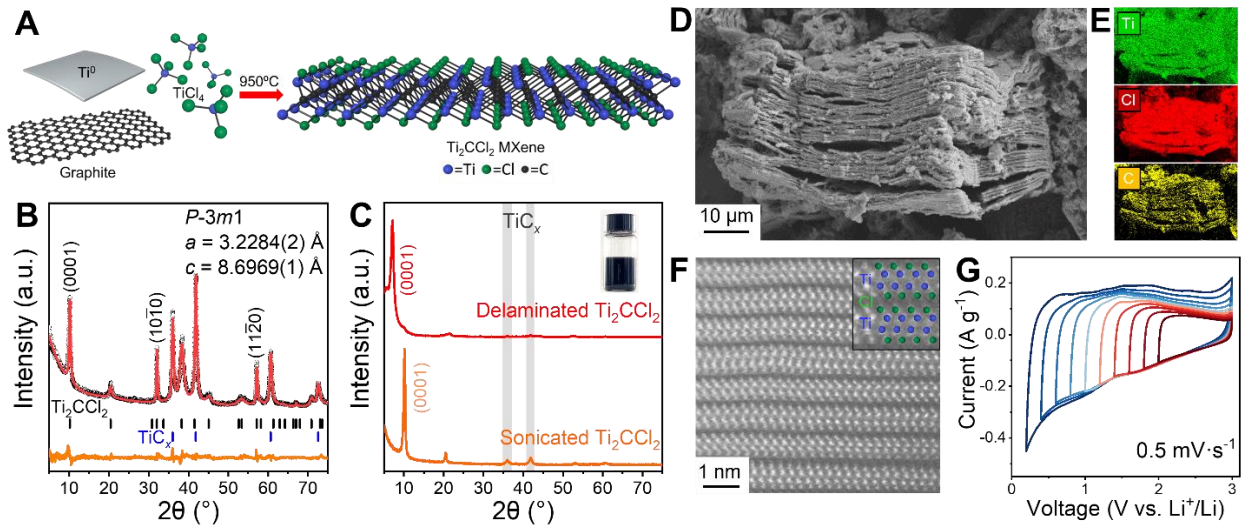

**Fig. 1. Direct synthesis and characterization of ds-Ti$_2$CCl$_2$ MXene**. (**A**) Schematic diagram of the synthesis. (**B**) XRD pattern and Rietveld refinement of ds-Ti$_2$CCl$_2$ prepared by reacting Ti, graphite and TiCl$_4$ at 950°C. (**C**) XRD patterns of dispersible delaminated and sonicated ds-Ti$_2$CCl$_2$ MXenes. Inset: Colloidal solution of the delaminated ds-Ti$_2$CCl$_2$. (**D**) SEM image and (**E**) EDX elemental mapping of a ds-Ti$_2$CCl$_2$ stack. (**F**) High resolution HAADF image representing the layered structure of ds-Ti$_2$CCl$_2$. (**G**) CV profiles of delaminated ds-Ti$_2$CCl$_2$ with various negative cut-off potentials.

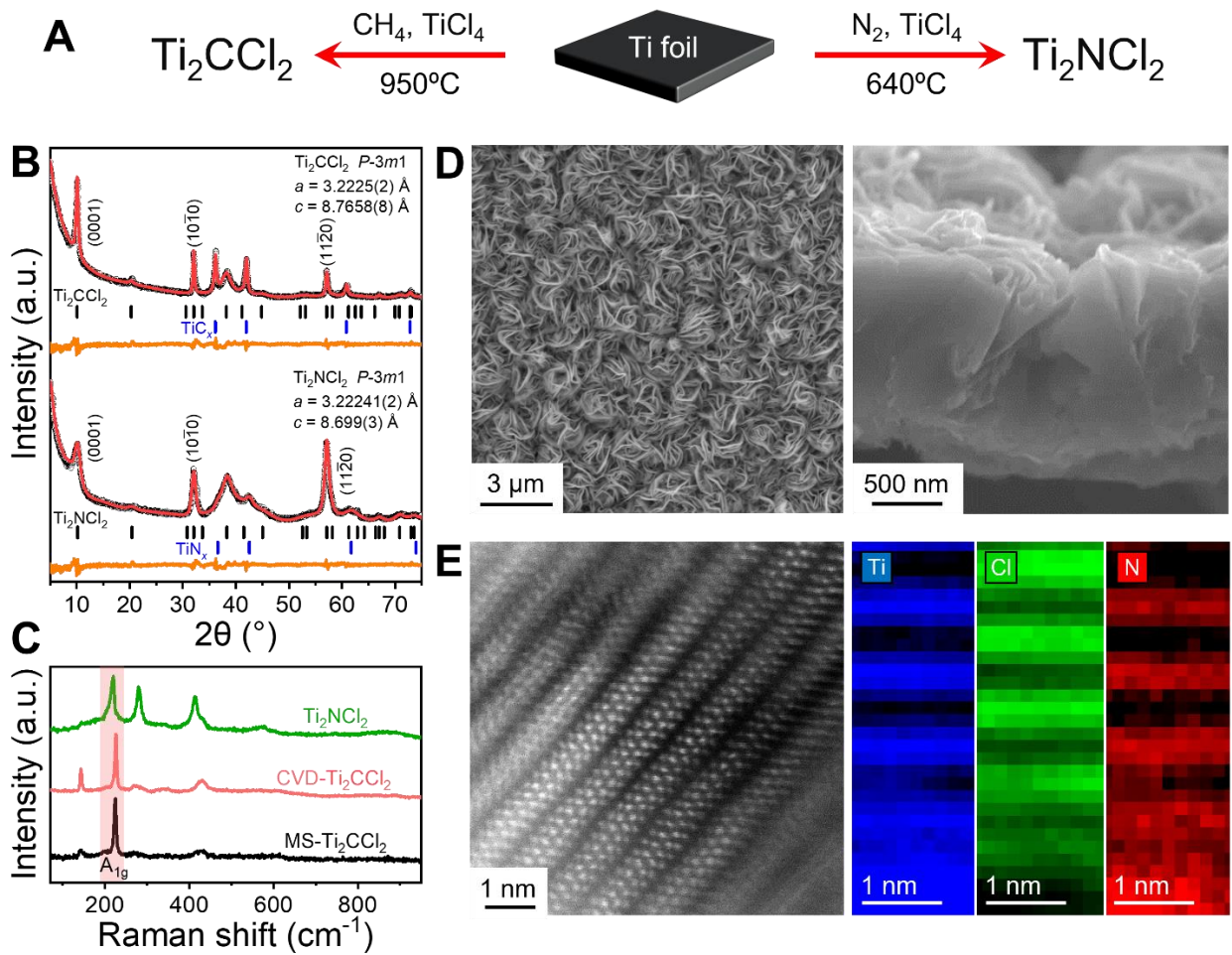

**Fig. 2. CVD growth of MXenes.** (**A**) Schematic diagram of the CVD reactions. (**B**) XRD patterns and Rietveld refinement for CVD-Ti$_2$CCl$_2$ and CVD-Ti$_2$NCl$_2$. (**C**) Raman spectra of CVD-Ti$_2$CCl$_2$ and CVD-Ti$_2$NCl$_2$ MXenes in comparison to that of a traditional MS-Ti$_2$CCl$_2$ MXene, which was synthesized by etching Ti$_2$AlC MAX phase with CdCl$_2$ molten salt. (**D**) Frontal and cross-sectional SEM images of CVD-Ti$_2$CCl$_2$. (**E**) High resolution HAADF images and EELS elemental mapping of CVD-Ti$_2$NCl$_2$.

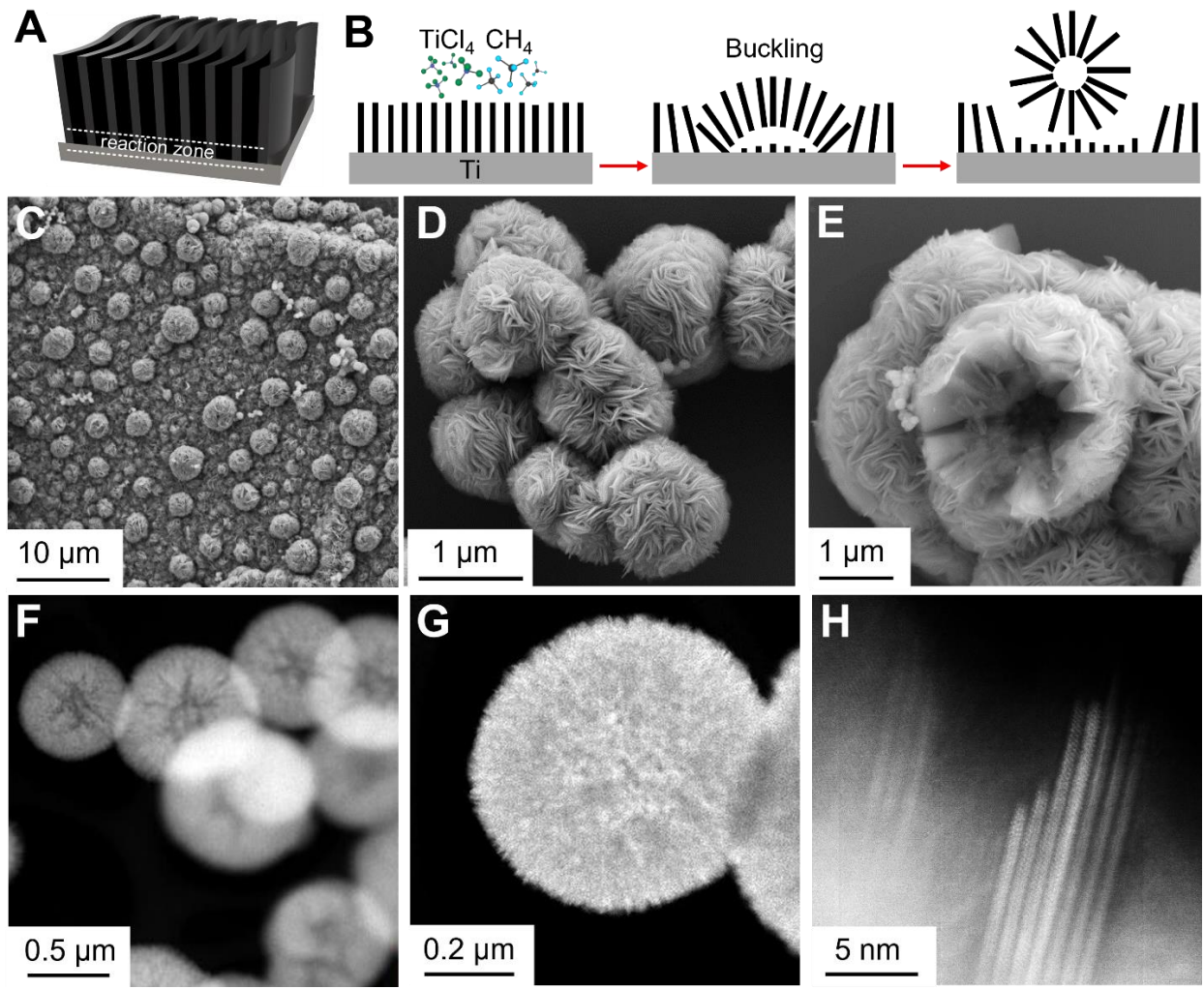

**Fig. 3. Morphologies of CVD-Ti$_2$CCl$_2$.** Schematic diagram illustrating the (**A**) reaction zone and (**B**) proposed buckling mechanism of CVD-Ti$_2$CCl$_2$ through which microspheres are formed. SEM images show that morphology of CVD-Ti$_2$CCl$_2$ can be varied by tuning reaction conditions: (**C**) Microspheres growing on carpets, (**D**) individual microspheres, and (**E**) a fragmented microsphere showing a hollow center. (**F, G, H**) STEM analysis further proves that vertically aligned MXene sheets constitute the microspheres, while a void is left at the center.